\documentclass{osa-article}
\usepackage{float}
\usepackage{commath}

\journal{oe}

\articletype{Research Article}

\begin{document}

\title{Performance limits of astronomical arrayed waveguide gratings on silica platform}

\author{Andreas Stoll,\authormark{1,*} Kalaga Madhav,\authormark{1} and Martin Roth\authormark{1}}

\address{\authormark{1}Leibniz-Institut f\"ur Astrophysik (AIP), An der Sternwarte 16, 14482 Potsdam, Germany}

\email{\authormark{*}astoll@aip.de} 


\flushbottom
\widowpenalty10000
\clubpenalty10000

\begin{abstract}
We present a numerical and experimental study of the impact of phase errors on the performance of large, high-resolution Arrayed Waveguide Gratings (AWG) for applications in astronomy. We use a scalar diffraction model to study the transmission spectrum of an AWG under random variations of the optical waveguide lengths. We simulate phase error correction by numerically trimming the lengths of the optical waveguides to the nearest integer multiple of the central wavelength. The optical length error distribution of a custom-fabricated silica AWG is measured using frequency-domain interferometry and Monte-Carlo fitting of interferogram intensities. In the end, we give an estimate for the phase-error limited size of a waveguide array manufactured using state-of-the-art technology. We show that post-processing eliminates phase errors as a performance limiting factor for astronomical spectroscopy in the H-band.
\end{abstract}

\section{Introduction}
\begin{figure}[h]
  \includegraphics[width=\linewidth]{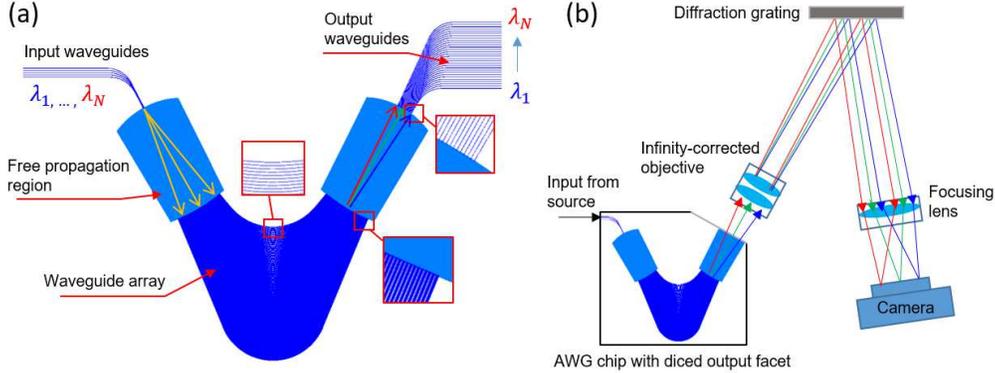}
  \caption{(a) Structure of a typical AWG device consisting of input waveguides, output waveguides, free propagation slab waveguides and the waveguide array. (b) Conceptual layout of astronomical AWG-based spectrograph. The free-space optics section performs two functions: (1) magnified re-imaging the AWG output spectrum onto the detector, (2) separating AWG spectral orders by cross-dispersion through diffraction grating, perpendicular to the plane of AWG dispersion.}
  \label{fig:fig_1}
\end{figure}

The demand for high-efficiency, high-resolution integrated spectrographs in astronomy has shifted Arrayed Waveguide Gratings (AWGs) into the focus of astrophotonics research. 
Arrayed waveguide gratings are planar integrated photonic waveguide structures which implement the functionality of an optical diffraction grating using the phased-array principle in the infrared and optical spectral range \cite{Smit88}. Figure \ref{fig:fig_1}(a) shows the typical layout of an AWG waveguide structure. The device consists of equidistant input waveguides/output waveguides for signal injection and harvesting, two slab waveguides called free propagation regions (FPR) which couple the signal in and out of the waveguide array centred between the FPRs. The waveguide array consists of many single mode waveguides obeying the geometrical rule $L_j = L_{j-1}+\Delta L$, where $L_j$ is the length of the $j^{th}$ waveguide. The constant nearest-neighbour path length difference (PLD) is defined as $\Delta L = m\lambda_0/n_{\text{eff}}$, where $m$ is the grating order, $\lambda_0$ is the central wavelength of operation, and $n_{\text{eff}}$ is the effective index of the array waveguides. As optical signals propagate through the multiple paths of the waveguide array, the phase difference between nearest neighbour paths after propagation varies linearly with the wavelength of the signal and is precisely zero for the central wavelength $\lambda_0$. The phase shift of the optical paths adds a wavelength-dependent tilt to the wavefronts emerging from the waveguide array on the output side of the AWG, creating a wavelength-dispersed focused image of the input at the location of the output waveguides. Since the nearest-neighbour phase shift cycles between $-\pi$ and $\pi$, the relation between wavelength and focused image location is only unique within a free spectral range (FSR), given by $FSR=\lambda_0/m$.
Arrayed waveguide gratings have been traditionally utilized as passive dense wavelength division multiplexers (DWDMs) in optical telecommunication networks. The use of AWGs as compact spectrographs in astronomy was first proposed by Bland-Hawthorn et al. \cite{Bland-Hawthorn06}.
Commercial telecom-grade AWG devices have been successfully repurposed for spectroscopic applications in astronomy \cite{Cvetojevic:09, Cvetojevic:12}.
The most prominent difference between AWG-based astronomical spectrographs and telecom-type AWGs is the absence of output waveguides. The requirement of high spectral resolution and instrument throughput suggests direct imaging of the output intensity distribution with a high-resolution camera, making the notion of discrete wavelength channels obsolete in the context. A conceptual sketch of an AWG-based spectrograph is shown in Figure \ref{fig:fig_1}(b). The output of the AWG is projected onto a camera via a bulk-optical system consisting of collimating and focusing optics as well as a blazed diffraction grating. Due to the relatively small FSR of the AWG and its cyclic behaviour causing overlapping of spectral orders, the blazed grating separates the spectral orders, allowing to cover a total wavelength range of several hundreds of nm. Note that cross-dispersion happens perpendicular to the plane of the drawing in Fig. 1 (b). Similar to astronomical echelle spectrographs, a two-dimensional image sensor is needed.

The requirements on spectral resolution and bandwidth, as well as the need for additional elements, such as on-chip calibration sources, call for specialized AWG designs, which push state-of-the art fabrication technology to its limits. The performance of an AWG critically depends on the correct phasing of the waveguide array, requiring a high accuracy of the lithographic fabrication process. Random spatial variations of material refractive index, core layer thickness and waveguide width translate into phase errors of a magnitude proportional to the mean waveguide array length, which impose a limit on the size of a functional AWG and therefore the maximum achievable spectral resolution. In a previous work, we have numerically investigated the impact of random spatial effective index variations on the performance of AWGs of various sizes, finding a critical threshold of $\sim 10^{-5}$ for the tolerable effective index variation at high spectral resolutions $\lambda/\Delta\lambda \geq 15000$ \cite{Stoll:17}. Phase error reduction techniques provide limited control over random phase deviations by inducing small changes of the waveguide effective index through UV-trimming \cite{Zauner:98}, thermo-optic \cite{Gehl:17} or electro-optic effects \cite{Jiang2008}. In the case of silica-on-silicon (SoS) and other passive material platforms, the choice of methods is restricted to UV-trimming and thermal tuning due to weak non-linearity and lack of electrical conductivity. 

The prospect of high-resolution AWG spectrographs using increasingly larger waveguide arrays in terms of waveguide number, average waveguide length and device foot-print raises concerns over the impact of fabrication related waveguide effective index variations. The effective index and the related group index $n_g=n_{\text{eff}}-\lambda dn_{\text{eff}}/d\lambda$ of an optical waveguide are sensitive to non-uniformities of the material refractive index as well as variations of the core width and thickness. Typical waveguide effective index fluctuations in AWGs depend on the material platform and range from $\delta n_{\text{eff}}\sim 10^{-4}$ in silicon AWGs to $\delta n_{\text{eff}}\sim 10^{-6}$ in low-refractive-index-contrast silica AWGs \cite{Okamoto:2014}. Taking the worst case $\delta n_{\text{eff}}\sim 10^{-4}$ as an example, we expect the optical length variation of a 15 mm long waveguide to exceed one wavelength, implying a phase error $>2\pi$, at 1550 nm.
In this work, we investigate the lower limit of feasibility for the implementation of a large-footprint, high-resolution AWG device on silica platform under the premise of reliable and well-controllable phase error trimming in a range $\leq 2\pi$. We numerically simulate the effects of very large random variations of the optical waveguide length on the order of several wavelengths and demonstrate the impact of path length error trimming to the nearest integer multiple of the central wavelength $\lambda_0$. Furthermore, we derive upper estimates of waveguide effective index variation from phase error measurements on a custom-designed SoS AWG device. In conclusion, we present an estimate of the maximum feasible AWG size and spectral resolution on a low index-contrast, SoS platform based on the phase accuracy of state-of-the-art manufacturing technology and spectral bandwidth requirements of astronomical spectroscopy applications.

\section{Methods}
\subsection{Numerical simulation of AWG transmission spectrum}
The theoretical AWG transmission spectrum is calculated in this work using a semi-analytical model based on scalar Fraunhofer diffraction \cite{Okamoto:2006}. Fundamental building blocks of the AWG, such as channel waveguides, slab waveguides and circular bends, are simulated using the beam propagation method (BPM). The results of the BPM simulations are then used to set up the semi-analytical transmission spectrum calculation. Phase errors are incorporated into the simulation by Gaussian randomization of the waveguide effective index across the waveguide array, resulting in random excess optical path lengths (OPLs). Transmission spectra are simulated in a wide wavelength interval 1300 nm - 1700 nm in order to detect long-range effects of small OPL errors. Simulations are performed for the cases of continuous  random OPL error distributions, as well as discrete distributions across integer multiples of the central wavelength, where the latter case is of significant relevance to the estimation of performance limits of a phase-error corrected AWG device.

\subsection{Optical path length error measurement}
Measurements of OPL deviations from design values in AWG devices were conducted by means of frequency-domain interferometry using a fibre-based Mach-Zehnder-Interferometer (MZI)  \cite{Takada:00, Takada:06}. A combination of a T100S-HP tunable laser and a CT440 component tester was used to obtain Mach-Zehnder interferograms of a device-under-test (DUT) in the wavelength range \linebreak 1500 nm - 1680 nm. A more detailed description of the experimental setup is given in section \ref{sec:mzi}. Phase error extraction is accomplished using the method of Monte-Carlo fitting of interference intensities \cite{Oh:12, Lim:13}, wherein the parameters of an AWG model are randomly varied in a gradient descent process, stochastically minimizing the residual sum of squares between simulated and measured transmission spectrum. A description of the interferogram fitting and results are given in section \ref{sec:mc-fit}.

\section{Numerical simulation of discrete path length errors in AWG devices}

\begin{figure}[H]
\centering
  \includegraphics[width=315pt]{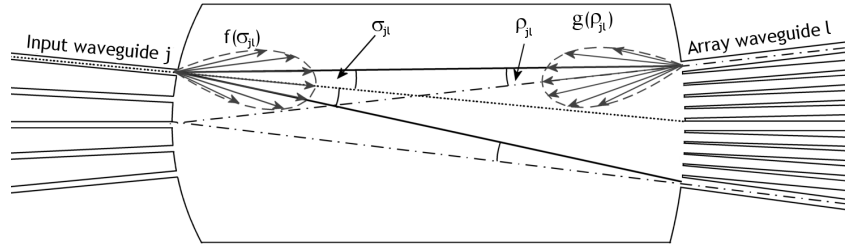}
  \caption{Principal sketch of a free propagation region with input wave\-guides and array wave\-guides shown.}
  \label{fig:FPR_IN}
\end{figure}

Simulations of the AWG transmission spectrum in the presence of random OPL errors were conducted using a semi-analytical 2D scalar model based on Fraunhofer diffraction in combination with the 3D BPM method, implemented in the commercial software RSoft BeamPROP. Fundamental properties of individual components of the AWG, such as channel waveguides and slab waveguides, were simulated using BPM. The relevant results to be obtained from BPM simulations were mode field distributions of the input waveguide, array wave\-guides, and FPR slab wave\-guides, as well as their wavelength-dependent effective indices. Following processing of the BPM results into an appropriate format, the data was used to initialize the scalar AWG model. The model works by calculating the AWG transmission function between input $j$ and output $k$ \cite{Okamoto:2006}:

\begin{subequations}

\begin{equation}\label{eq:AWG_analytical}
T_{jk}(\lambda)=\abs{\sum_{l=1}^{M} c_{jk}(l)\exp\left(-i\Omega_{jk}(l)\right)}^2
\end{equation}

\begin{equation}
c_{jk}(l) = f(\sigma_{jl})g(\rho_{jl})g(\rho_{kl})h(\sigma_{kl})
\end{equation}

\begin{equation}
\Omega_{jk}(l) = \frac{2\pi}{\lambda}\left( n_s(r_{jl}+r_{kl})+ n_a(l\Delta L+L_0) \right),
\end{equation}
\end{subequations}
where $f$, $g$ and $h$ are emittance/acceptance functions of the input wave\-guides, array wave\-guides and output wave\-guides, respectively, $n_s$ and $n_a$ are the slab waveguide and array waveguide effective indices, $r_{jl}$ and $r_{jk}$ are the distances between array waveguide $l$ and input $j$ / output $k$, respectively, and $\Delta L$ is the waveguide array path length increment. The waveguide emittance functions and effective indices are obtained from BPM simulations.
Figure \ref{fig:FPR_IN} shows a sketch of the FPR geometry, as assumed in the semi-analytical model. Only the input side of the AWG is shown since the output side is identical to the input due to the symmetry of the AWG device.

The semi-analytical method offers the advantage of great time-efficiency compared to a BPM simulation, especially in the case of large AWGs which may take several hours to simulate using BPM. In contrast, the runtime of a semi-analytical calculation of the transmission function is on the order of a few minutes, not counting the preliminary BPM simulations which have to be performed only once for any particular AWG design. Comparisons of semi-analytical AWG transmission spectra with results obtained from direct BPM simulations show a good agreement between the two methods.

\begin{figure}[H]
  \includegraphics[width=\linewidth]{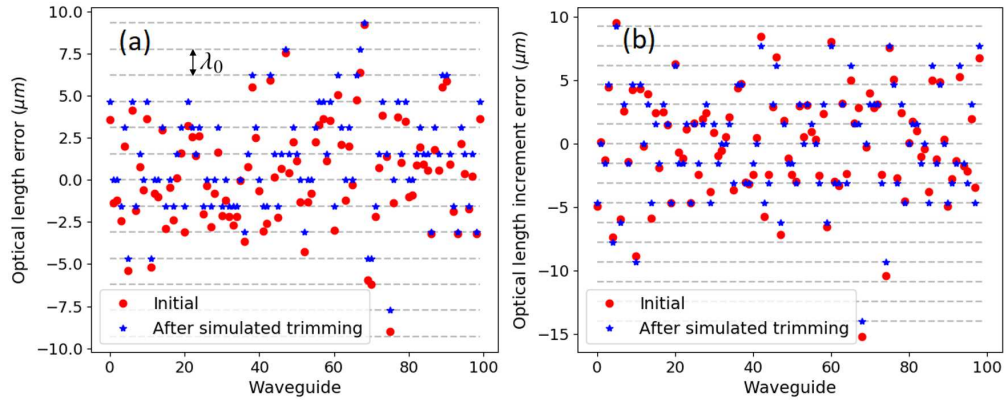}
  \caption{(a) Randomly generated distribution of OPL errors before and after simulated trimming, (b) Distribution of associated waveguide array path length increment errors}
  \label{fig:fig1}
\end{figure}

For the simulation of the effects of very large path length errors on the transmission spectrum of an AWG, we have used a small AWG model consisting of 100 array wave\-guides. The material platform was assumed to be SoS, however, the method described in this work is applicable to any material system. A waveguide structure defined by a 3.4 $\mu$m thick core layer with a refractive index of 1.473 (at 1550 nm) surrounded by a cladding with refractive index 1.444 (at 1550 nm) was assumed in the BPM simulations. Optical path length errors were included into the model by allowing the waveguide effective index $n_a$ in Eq. \eqref{eq:AWG_analytical} to vary across the waveguide array and adding a Gaussian randomization with a defined standard deviation to the values of $n_a$. In order to demonstrate and compare the impact of very large path length errors with continuous and discrete distribution, the standard deviation $\sigma$ of the random effective index distribution across the wave\-guides was varied between $10^{-5}$ and $3\times 10^{-3}$. A path length increment of 100 $\mu$m was set for the waveguide array. With an average waveguide length of 15 mm, the effective index variation translates to OPL errors with standard deviations of up to 15 $\mu$m. Although path length errors of this magnitude are not realistic in most practical cases, they serve to demonstrate the effect of a discretization of the path length error distribution for different magnitudes of random effective index variation. Waveguide array trimming was simulated by lifting the optical length errors on each waveguide to the nearest upper integer multiple of the central wavelength $\lambda_0$.
Figure \ref{fig:fig1} shows OPL errors for a randomization of the effective index error with standard deviation $\sigma=3\times 10^{-4}$. The excess optical lengths are varying between -6 $\mu$m and 4 $\mu$m. Simulated trimming arranges the length deviations in discrete bands with spacing $\lambda_0$, marked by grey dashed lines, centred around zero (Figure \ref{fig:fig1}(a)). 
Figure \ref{fig:fig1}(b) shows the error of the nearest-neighbour PLD, given by the discrete derivative of the optical length error distribution in Figure \ref{fig:fig1}(a).
\begin{figure}[H]
  \includegraphics[width=\linewidth]{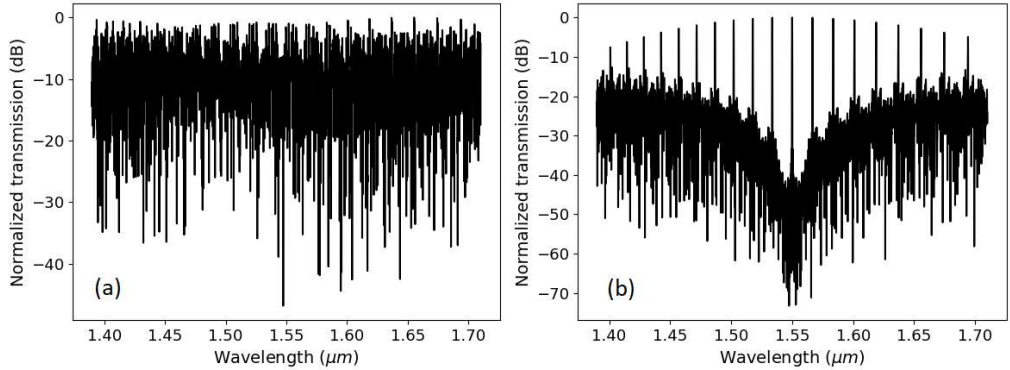}
  \caption{Transmission spectrum of the simulated AWG in presence of randomized OPL errors with $\sigma=3\times 10^{-4}$. (a) Spectrum without trimming and (b) after simulated trimming of optical lengths to nearest integer multiples of the central wavelength}
  \label{fig:fig2}
\end{figure}
Figure \ref{fig:fig2}(a) shows the calculated transmission function \eqref{eq:AWG_analytical} between the central input and central output channel of the simulated AWG, assuming the path length error distribution shown in Figure \ref{fig:fig1} before simulated trimming is applied. The resulting OPL errors exceed $\lambda_0$, hence the channel transmission peak is completely obscured by crosstalk, rendering the AWG device unusable. The crosstalk level near the central wavelength decreases significantly after simulated trimming despite residual discretely distributed length errors, as shown in Figure \ref{fig:fig2}(b). The phase-error induced crosstalk component becomes zero at the central wavelength $\lambda_0$ and increases further away from $\lambda_0$. The total crosstalk level varies from -50 dB at 1.55 $\mu$m to -20 dB at 1.4 $\mu$m and 1.7 $\mu$m. The increase of crosstalk at off-centre wavelengths is affected by the magnitude of effective index errors. Figure \ref{fig:fig3} shows a comparison between transmission spectra for effective index standard deviations $\sigma=3 \times 10^{-3}$ (Figure \ref{fig:fig3}(a)) and $\sigma=10^{-5}$ (Figure \ref{fig:fig3}(b)) after simulated trimming. The usable wavelength range shrinks to 64 nm in the case of $\sigma=3\times 10^{-3}$, while the entire simulation range of 300 nm is usable at $\sigma=10^{-5}$. The reduction of crosstalk due to optical length trimming is observed on all wavelength channels of the AWG, as shown in Figure \ref{fig:AWG_allchans}.
\vspace{\baselineskip}
\begin{figure}[H]
  \includegraphics[width=\linewidth]{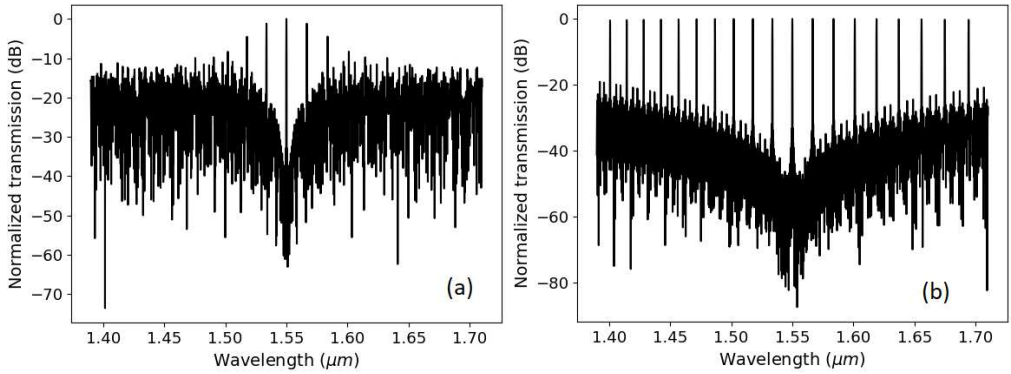}
  \caption{Comparison of wide-range transmission after simulated trimming of phase errors: (a) $\sigma=3\times 10^{-3}$ and (b) $\sigma=10^{-5}$. Larger phase errors significantly reduce the usable wavelength range.}
  \label{fig:fig3}
\end{figure}
\begin{figure}[H]
  \centering
  \includegraphics[width= \linewidth]{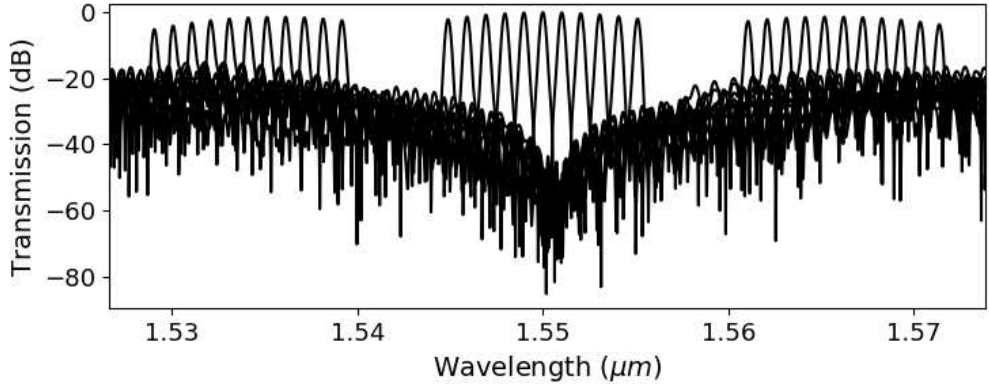}
  \caption{Transmission spectrum of all 11 wavelength channels of the simulated AWG showing main spectral order around the central wavelength 1550 nm and two adjacent spectral orders at 1550 nm $\pm$ 16 nm. Waveguide effective index errors with $\sigma=3\times 10^{-3}$ used for demonstration. }
  \label{fig:AWG_allchans}
\end{figure}
The total usable wavelength range can be estimated by determining the neighbourhood of $\lambda_0$, inside which the absolute values of phase errors induced by the discrete excess optical lengths do not exceed $\pi$. For a transmission from central input to central output, function \eqref{eq:AWG_analytical} simplifies to
\begin{equation}\label{eq:eq1}
	T(\lambda)=\abs{\sum_{k=1}^{N}a_k\exp\left(-i \frac{2\pi}{\lambda} n_{a,k} L_{k}\right)}^2,
\end{equation}
where $N$ is the number of wave\-guides in the array, $a_k$ is the amplitude of the $k^{th}$ contribution to the total field coupled into the output waveguide, $n_{a,k}$ and $L_k$ are the effective index and geometrical length of the $k^{th}$ waveguide, respectively.
To obtain a safe upper boundary for the effective index variability, we make the maximal assumption of negligible geometrical path length errors, hence $\delta L_{k}=0$. In the phase term, we set $n_{a,k}={n_a}+\delta n_{a,k}$, where $\delta n_{a,k}$ is the deviation of the waveguide effective index from the design value, and isolate the wavelength-dependent nearest-neighbour phase error relative to the central wavelength $\lambda_0$:
\begin{equation}\label{eq:eq2}
	\delta\phi_{k}(\lambda)=2\pi\left(\frac{1}{\lambda}-\frac{1}{\lambda_0}\right)\left(\delta n_{a,{k+1}} L_{k+1}-\delta n_{a,k} L_{k}\right) \approx 2\pi\left(\frac{1}{\lambda}-\frac{1}{\lambda_0}\right)L_k\left(\delta n_{a,{k+1}} -\delta n_{a,k}\right).
\end{equation}
Next, we estimate the upper limit of the array waveguide length using the standard deviation $\sigma_{\delta\phi_{k}}$ of the relative phase error
\begin{equation}\label{eq:eq2_1}
	\sigma_{\delta\phi_{k}}=2\pi\abs{\frac{1}{\lambda}-\frac{1}{\lambda_0}}L_k\sqrt{2}\sigma_{n_a},
\end{equation}
where $\sigma_{n_a}$ is the standard deviation of the waveguide effective index.
Taking the long edge of the waveguide array at $k=N$ as reference under the condition $\sigma_{\delta\phi_{k}}\leq\pi$, we obtain
\begin{equation}\label{eq:eq3}
	\abs{\lambda-\lambda_0} \leq \frac{\lambda_0\lambda}{\sqrt{8}\sigma_{n_a} L_N}.
\end{equation}
The choice $k=N$ does not take into account that wave\-guides near the edge of the array typically carry a relatively small fraction of the total power. However, since the power distribution can be arbitrary in general, we obtain a safe estimate by assuming the longest waveguide as reference.
The left hand side of Eq. \eqref{eq:eq3} can be interpreted as one half of the free spectral range of the effective-index-variation induced excess optical lengths. On the right hand side, we can approximate $\lambda\approx \lambda_0$, since $\abs{\lambda-\lambda_0}\ll \lambda_0$. In order to estimate the usable wavelength range of a physical AWG, a reasonable experimental constraint for $\sigma_{n_a}$ must be obtained from OPL error measurements on a fabricated AWG device. 

\section{Custom-fabricated Silica-on-Silicon AWG}
\begin{figure}[H]
  \includegraphics[width=\linewidth]{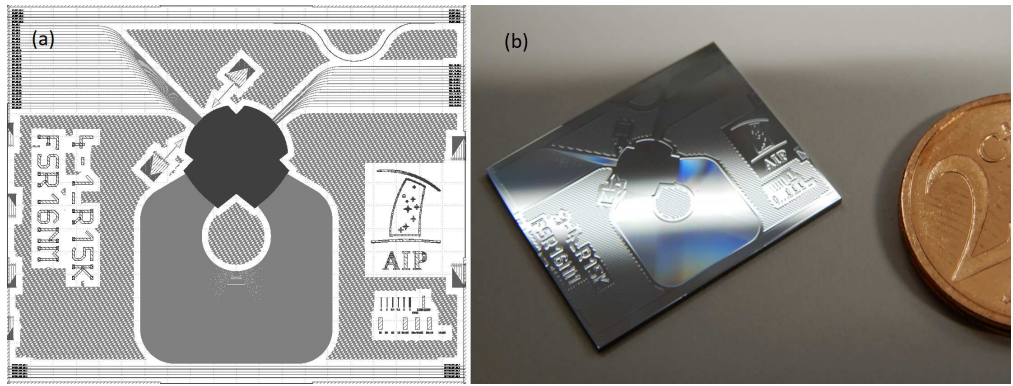}
  \caption{Custom-designed and fabricated AWG device on silica platform. (a) Lithographic mask layout. (b) Fabricated AWG chip}
  \label{fig:fig4}
\end{figure}

\begin{figure}[H]
  \includegraphics[width=\linewidth]{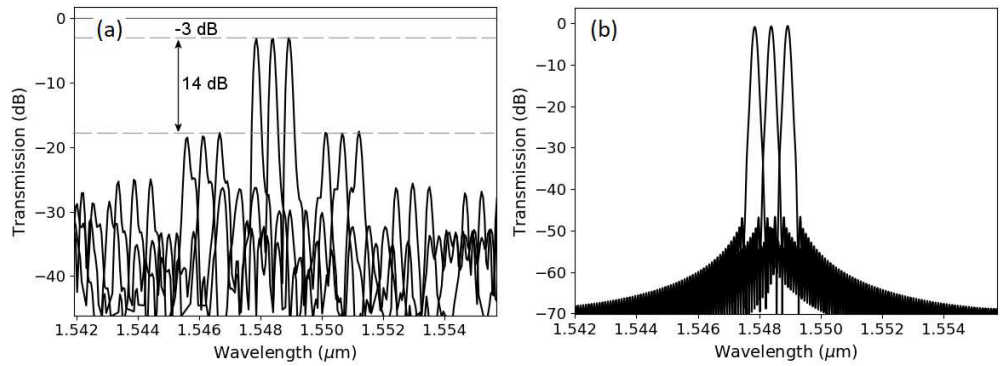}
  \caption{(a) Measured power transmission curves of three selected wavelength channels. (b) Simulated ideal power transmission curves.}
  \label{fig:AWG_measured}
\end{figure}

The AWG device used in the experiment was designed in-house and fabricated by an external foundry (Enablence USA Components, Inc.). The waveguide structure was designed and optimized for operation in the astronomical H-band in the wavelength range 1500 nm - 1800 nm. The theoretical FSR and target spectral resolution were defined as 16 nm and 0.1 nm, respectively. Due to the requirement of low device loss and high refractive index uniformity, a SoS material platform with a refractive index contrast of 2$\%$ was chosen for the design, as the fabrication methods of SoS technology are well-developed and reliable. In order to guarantee single mode operation of the waveguide array in the entire wavelength range, the single-mode cut-off cross-section of the wave\-guides was determined by BPM simulation to be 3.4 $\mu$m $\times$ 3.4 $\mu$m at 1500 nm. Therefore, a 3.4 $\mu$m thick guiding layer was assumed for the entire device. Buffer and cladding layers were assumed to be silica with a refractive index of 1.444 at 1550 nm, suggesting a required core refractive index of 1.474. The radius of waveguide curvature was limited down to 1.5 mm in order to minimize losses due to bend leakage. Numerical analysis of evanescent coupling between adjacent array wave\-guides has shown that a minimum distance of 10 $\mu$m - 11$\mu$m between array wave\-guides at the array-FPR interface must be observed in order to avoid excessive coupling between nearest neighbours. A minimum distance of 10.57 $\mu$m between the wave\-guides at the array-FPR interface was chosen based on the results. Given the FSR of 16 nm and a grating period of 10.57 $\mu$m, a focal length of 4.5 mm was chosen to meet the requirement of spectral resolution. The aperture width of the waveguide grating was designed to limit beam truncation losses to 1$\%$ of the total input power while keeping the foot-print of the AWG structure reasonably small. The waveguide array, consisting of 179 wave\-guides with a nearest-neighbour PLD of 99.98 $\mu$m, was arranged in a folded geometry to keep the AWG small while at the same time providing sufficient space for waveguide length accumulation. The central wavelength $\lambda_0$ was defined as 1550 nm. Structurally, the waveguide array consists of alternating straight segments and curved segments with a curvature radius of 1.5 mm. Coupling of signal in and out of the AWG is realized by two arrays of input/output wave\-guides on each side of the AWG. Since the AWG is intended to be used unidirectionally, the design incorporates 15 input wave\-guides and 30 output wave\-guides arranged with a spacing of 15 $\mu$m. The output wave\-guides are required for characterization and will be removed by dicing at the FPR-output wave\-guide interface and polishing the facet in preparation for use in a spectrograph. The final lithographic mask layout of the AWG is shown in Figure \ref{fig:fig4}(a). Figure \ref{fig:fig4}(b) shows the fabricated AWG chip.
The design was fabricated on a 6-inch Si-wafer using atmospheric pressure chemical vapour deposition (APCVD). A 3.4 $\mu$m thick Ge-doped silica core layer was deposited on top of a 20 $\mu$m thick thermal oxide buffer followed by etching of the waveguide structure and deposition of a 15 $\mu$m silica cladding.   
The power transmission curve of the fabricated AWG was measured using an optical wavelength meter (OWM) which was synchronized with a tunable laser source (TLS). The TLS wavelength was tuned in discrete 0.05 nm steps and the transmitted power level was recorded by a PC. Measurements were normalized with the transmission curve of a straight 20 mm long reference waveguide on the same chip. The transmission curves of three selected wavelength channels are shown in Figure \ref{fig:AWG_measured}(a). The transmission peaks at -3 dB on all three output channels. The maximum measured crosstalk is -17 dB. Figure \ref{fig:AWG_measured}(b) shows the theoretical transmission curves of the ideal AWG for comparison.

\section{Frequency-domain interferogram measurement}\label{sec:mzi}
\begin{figure}[H]
  \centering
  \includegraphics[width=300pt]{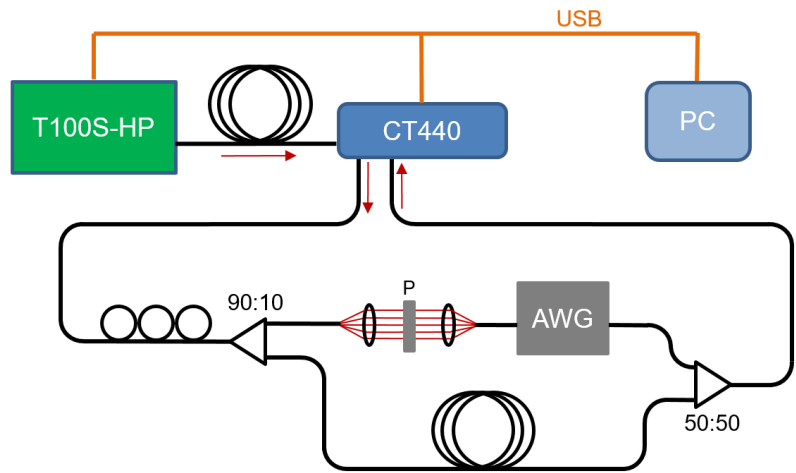}
  \caption{Schematic layout of the Mach-Zehnder frequency domain interferometer setup.}
  \label{fig:fig5}
\end{figure}

\begin{figure}[H]
  \includegraphics[width=\linewidth]{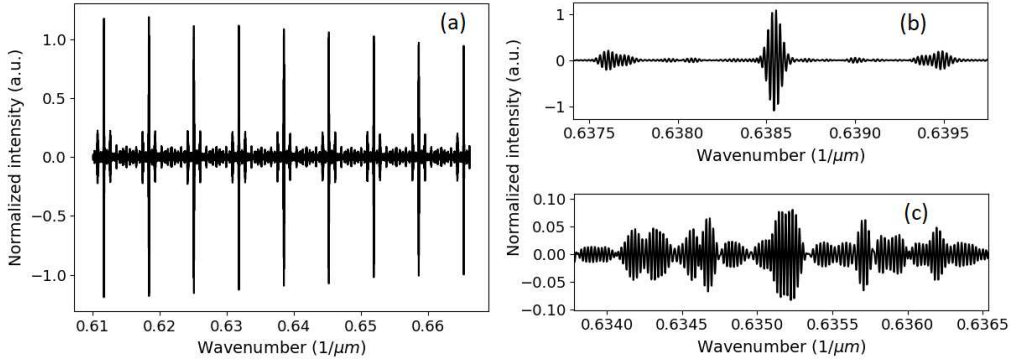}
  \caption{Normalized interferogram of the fabricated silica AWG. (a) Full wavelength range 1500 nm - 1640 nm, (b) Enlarged view of main transmission peak, (c) Enlarged view shows the intermediate crosstalk.}
  \label{fig:fig6}
\end{figure}

We have obtained the phase error distribution of the custom-fabricated AWG device using frequency-domain interferometry method with a fibre-based Mach-Zehnder configuration (Figure \ref{fig:fig5}). The interferometer consists of two Mach-Zehnder arms with an optical PLD of 8 cm, whereby the reference arm is longer than the DUT arm containing the AWG. Input power is divided between the AWG arm and reference arm with a ratio of 90:10, respectively. The AWG arm of the interferometer contains a rotating polariser in a free-space U-bench configuration. Polarization is controlled globally using a fibre-based polarization controller before the first power divider. After propagation through the interferometer the signals are combined using a 50:50 fibre combiner, resulting in a final power ratio of 45:5.
Swept-wavelength measurements were made using a combination of a T100S-HP tunable laser source and a CT440 optical component tester in the wavelength range 1500 nm - 1640 nm. TE-polarized interferograms were recorded with a resolution of 5 pm and a tuning speed of 30 nm/s. Background subtraction, fringe contrast normalization and dispersion correction were performed on the AWG interferogram using a calibration interferogram of a straight test waveguide on the same chip. The optical PLD between the AWG arm and reference arm of the interferometer was tuned to obtain interferogram oscillations in the PLD range 30000 $\mu$m - 58000 $\mu$m, which correspond directly to the optical length differences of the MZI paths, while avoiding overlapping of the MZI interferogram spectrum with the self-interference spectrum of the AWG. Figure \ref{fig:fig6}(a) shows the AWG interferogram function after normalization and dispersion correction. The measurement range covers 8 spectral orders of the AWG. Figures \ref{fig:fig6}(b) and \ref{fig:fig6}(c) show a close-up of the main AWG channel transmission peak and the phase error induced crosstalk between the main spectral peaks, respectively.

\section{Interferogram analysis by Monte-Carlo fitting of interference intensities}\label{sec:mc-fit}
\begin{figure}[H]
  \includegraphics[width=\linewidth]{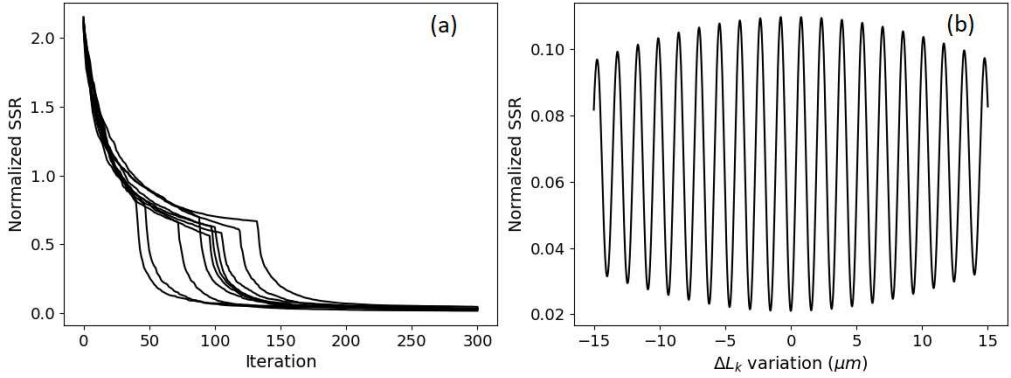}
  \caption{(a) Evolution of the SSR cost function over 300 iterations of the fitting algorithm. (b) 1D slice of the cost function under variation of the optical length of a single waveguide.}
  \label{fig:fig7}
\end{figure}

\begin{figure}[H]
  \includegraphics[width=\linewidth]{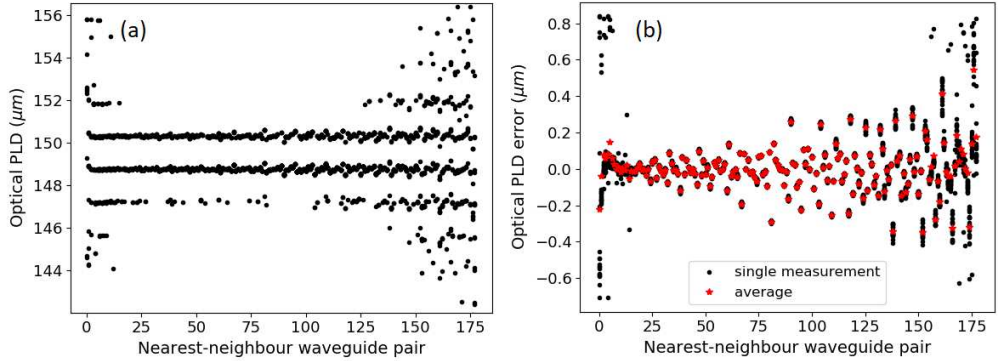}
  \caption{Fitting results after convergence: (a) Absolute optical nearest-neighbour PLD, (b) Relative optical PLD error. Values ordered from shortest to longest waveguide.}
  \label{fig:fig8}
\end{figure}

We have analysed the measured interferograms using the method of Monte-Carlo fitting of interference intensities. After normalizing and removing the low frequency components, the interferogram can be expressed as \cite{Oh:12}
\begin{equation}\label{eq:eq4}
	I(\nu)=\sum_{k=1}^{N} c_k\cos\left(2\pi\nu \Delta L_k\right),
\end{equation}
where $\nu=1/\lambda$ is the wave-number, $c_k$ is the amplitude of the $k^{th}$ interferogram component and $\Delta L_k$ is the optical PLD between the $k^{th}$ path of the AWG arm and the reference path of the MZI. An interferogram resulting from the interference of $N$ optical paths is defined by 2N parameters, as each component of Eq. \eqref{eq:eq4} is defined by its amplitude $c_k$ and interference fringe period $1/\Delta L_k$. The fitting process is driven by small random variations of $c_k$ and $\Delta L_k$ under minimization of the sum of squared residuals (SSR) between Eq. \eqref{eq:eq4} and the experimental data. In order to reduce convergence time, we obtained initial values for $\Delta L_k$ from the Fourier spectrum of the measured interferogram and initialized $c_k$ with a Gaussian profile approximating the actual amplitude distribution. The ranges of random parameter variation were chosen to be 1.6 $\mu$m and $10^{-3}$ for $\Delta L_k$ and $c_k$, respectively. Fitting was performed on a truncated interferogram in the wavelength range 1520 nm - 1590 nm in order to reduce errors originating from random thermal drift of the MZI during the measurement. Rapid convergence was observed after 40 - 130 iterations through 358 fit parameters. Figure \ref{fig:fig7}(a) shows the SSR of 10 fitting runs with a series of independent measurements. Each run was terminated after 600 iterations, of which 300 are shown in the figure. The SSR curves show two distinct phases of convergence: The SSR data shows two distinct phases of convergence: initial gradient descent into a local minimum of the cost function followed by a rapid descent into the global minimum. Figure \ref{fig:fig7}(b) shows a 1-D slice of the 2N-dimensional SSR function under variation of the optical length of a single waveguide. The locations of the function minima are approximately $\lambda_0$ - periodic, as expected from Eq. \nolinebreak \eqref{eq:eq4}.\linebreak The measured runtime of a single-threaded Python implementation of the fitting algorithm performing 600 iterations was 180 s on a Core i5-8250U CPU running at 3.38 GHz.
The superimposed fitting results of 20 independent interferogram measurements are shown in Figure \ref{fig:fig8}. Figure \ref{fig:fig8}(a) shows the optical PLD between nearest-neighbour wave\-guides after convergence. The convergence of parameter values into discrete bands separated by the central wavelength $\lambda_0$ reflects the periodic structure of the cost function. The optical PLD values are scattered across multiple bands due to the limitation of absolute measurement accuracy by the narrow wavelength range of the fitted interferogram. However, a high relative phase measurement accuracy of $<0.1^{\circ}$ was reported in \cite{Lim:13}.
We obtained relative optical PLD errors by wrapping the scattered parameter values with periodicity $\lambda_0$ and subtracting the mean value. The final result of the optical PLD error measurement is shown in Figure \ref{fig:fig8}(b). The PLD error varies by $\pm$ 0.3 $\mu$m in the centre of the waveguide array. The short edge of the waveguide array shows smaller variation with $\pm$ 0.1 $\mu$m, while the variation on the long edge is approximately $\pm$ 0.4 $\mu$m. The standard deviation of the PLD error distribution is 0.123 $\mu$m. The measurement accuracy varies across the waveguide array between 250 nm at the edges and 5 nm in the central region of the array. 

\section{Extrapolated size limit for silica waveguide arrays}
\begin{figure}[H]\centering
  \includegraphics[width=250pt]{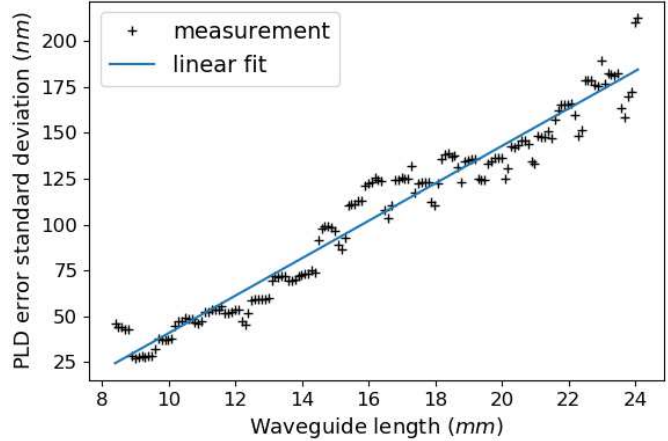}
  \caption{Measured standard deviation of the optical waveguide length errors in relation to the geometrical waveguide length. A linear proportionality between waveguide length and optical length variation is observed.}
  \label{fig:fig9}
\end{figure}

The fitting results described in the previous section indicate a direct proportionality between optical PLD errors and waveguide length. From this data, we have estimated the variation of the waveguide group index $n_g$ in the experimental AWG. We have calculated the average standard deviation of the optical PLD errors in a moving window containing 20 data points, finding a linear relation between waveguide length and nearest-neighbour PLD error. The slope of the linear fit of the PLD error in Figure \ref{fig:fig9} equals $\sqrt{2}\sigma_{n_g}$ and, neglecting a possible variation of the geometrical waveguide length, we obtain an upper estimate $\sigma_{n_g}=(7.2\pm 1.4)\times 10^{-6}$. 

For the estimate of the spectral resolution limit of a trimmed AWG device, we require a wavelength coverage of 300 nm in the astronomical H-band and retain the design parameters of the AWG under test, except for a linear scale-up of focal length and number of array wave\-guides. In Eq. \eqref{eq:eq3}, we assume $\sigma_{n_a}\approx\sigma_{n_g}$ and obtain the limit for the length of the longest array waveguide as $L_{N} \leq 7.86\times 10^{5}$\nolinebreak $\mu$m. Given the geometrical path length increment of 99.98 $\mu$m of the AWG, we calculate a waveguide array size limit of 7862 wave\-guides. Since the resolving power of an AWG is directly proportional to the waveguide array size, we can estimate a theoretical limit for the resolving power as  $R = 6.59\times 10^{5}$, or a spectral resolution of $\Delta\lambda = 2.35$ pm at 1550 nm. 

This remarkably high resolving power is calculated under the assumption that phase errors are the only limiting factor in the fabrication of AWGs with a large number of wave\-guides and large path length increments, leading to very long array wave\-guides. An AWG of the above size would cover an area of approximately 0.62 $m^2$, or 17.53 standard 6-inch wafers if implemented on a low-refractive-index SoS platform, which is theoretically feasible using a segmented fabrication strategy and post-fabrication assembly of the segments into a single structure, but highly impractical due to the extremely large foot-print. First and foremost, this result shows that the trimming of optical waveguide length errors to the nearest integer multiple of the central wavelength can facilitate the manufacturing of AWGs with almost arbitrarily long wave\-guides, limited only by the size of the wafer and the waveguide propagation loss. Post-processing can therefore effectively eliminate phase errors as a performance limiting factor. Phase error correction by UV trimming of AWG devices has been successfully demonstrated by Zauner et al. \cite{Zauner:98} using a 248 nm excimer laser on a hydrogen-loaded silica AWG. Silica photosensitivity enhancement methods based on hydrogen loading \cite{Lemaire:93} allow for induced refractive index changes on the order of $10^{-3}$, which exceeds the measured effective index variability in this work by roughly two orders of magnitude, ensuring the feasibility of UV trimming of effective index variations of the observed magnitude. However, hydrogen loading is a very time consuming process which requires exposure of the AWG chip to a high-pressure hydrogen atmosphere over several days prior to UV trimming. A high photosensitivity without the need for hydrogen loading was reported for waveguide structures fabricated by plasma-enhanced chemical vapour deposition (PECVD) and APCVD methods \cite{Nishii:96, Saito:98}, with refractive index tuning ranges between $10^{-3}$ and $10^{-2}$. If we assume the case of a 10 mm long waveguide, changing the optical length by a maximum amount $\lambda_0=$ 1.55 $\mu$m would require a material refractive index change of the core and surrounding cladding by $1.55\times 10^{-4}$ over the entire waveguide length. This requirement is gradually relaxed for longer wave\-guides, falling below $10^{-5}$ for wave\-guides lengths in excess of 155 mm.

\section{Conclusion}
In this work, we have investigated the implications of phase errors and phase error trimming for the application of large, high-resolution Arrayed Waveguide Gratings in astronomical spectroscopy. We have numerically studied the performance impact of random variations of optical waveguide length in excess of the central wavelength $\lambda_0$ using a scalar diffraction model of the AWG. We have shown that trimming of the optical waveguide length to the nearest integer multiple of $\lambda_0$ significantly improves AWG performance even if the optical length errors are scattered across a wide range and cannot be fully compensated. We have derived a simple lower estimate for a feasible waveguide array size on SoS platform using state-of-the-art manufacturing technology, based on the operating wavelength range requirement on the AWG and the magnitude of waveguide effective index variations. In order to obtain a realistic phase error distribution, we have used swept-wavelength frequency-domain interferometry to measure the phase errors of a custom-fabricated test AWG with 179 array wave\-guides, a theoretical spectral resolution of 0.1 nm, a free spectral range of 16 nm and a mean waveguide length of 12.5 mm. Interferogram analysis was performed using a Monte-Carlo interference fringe fitting algorithm with a series of interferograms recorded in the range 1520 nm - 1590 nm. The fitting results showed optical length variations between 0.1 $\mu$m and 0.4 $\mu$m across the waveguide array, with a tendency for larger variations at larger waveguide lengths. Using linear fitting of the resulting waveguide length dependent PLD variation distribution, we have obtained a standard deviation $\sigma_{n_{g}}=(7.2\pm 1.4)\times 10^{-6}$ for the group index variation of the array wave\-guides on the AWG chip. We extrapolated the optical waveguide length errors based on the measurement result and obtained an estimate for the phase-error limited waveguide array size of 7862 wave\-guides and a theoretical resolving power limit of $\lambda/\Delta\lambda=6.59 \times 10^{5}$. We have concluded from this estimate that the spectral resolution of a phase-error trimmed AWG is limited by the size of the wafer rather than by random OPL variations. In conclusion, this work shows the importance of post-processing in the manufacturing of high-performance, high-resolution AWG devices for astronomy, as well as the potential possibilities yet to be realized in practice. Future research in the field of AWG-based spectrographs must treat both phase-error measurement techniques and phase-error correcting post-processing as important elements in the development of high-resolution astronomy-grade AWG spectrographs.

\section*{Funding}
This work is supported by the BMBF project “Meta-ZIK Astrooptics” (grant No. 03Z22A511).

\section*{Acknowledgments}
This work is supported by the BMBF project “Meta-ZIK Astrooptics” (grant No. 03Z22A511).
We express our special thanks to Prof. Martin Schell, Fraunhofer HHI, Berlin, for fruitful discussions and useful advice.

\section*{Disclosures}
The authors declare no conflicts of interest.

\bibliography{manuscript}

\end{document}